# Charge sensing in enhancement mode double-top-gated metal-oxide-semiconductor quantum dots

E.P. Nordberg,[1,2] H.L. Stalford,[1,3] R. Young,[1] G.A. Ten Eyck,[1] K. Eng,[1] L.A. Tracy,[1] K.D. Childs,[1] J.R. Wendt,[1] R.K. Grubbs,[1] J. Stevens,[1] M.P. Lilly,[4] M.A. Eriksson,[2] and M.S. Carroll[1]

[1]Sandia National Laboratories, Albuquerque, New Mexico 87185
[2]University of Wisconsin-Madison, Madison, Wisconsin 53706
[3]University of Oklahoma, Norman, Oklahoma 73019
[4]Center for Integrated Nanotechnologies, Sandia National Laboratories, Albuquerque, New Mexico 87185

Abstract

Laterally coupled charge sensing of quantum dots is highly desirable, because it enables measurement even when conduction through the quantum dot itself is suppressed. In this work, we demonstrate such charge sensing in a double top gated MOS system. The current through a point contact constriction integrated near a quantum dot shows sharp 2% changes corresponding to charge transitions between the dot and a nearby lead. We extract the coupling capacitance between the charge sensor and the quantum dot, and we show that it agrees well with a 3D capacitance model of the integrated sensor and quantum dot system.

In recent years, semiconductor lateral quantum dots have emerged as an appealing approach to quantum computing.[1-4] The demonstration of electrically controlled spin qubits in GaAs/AlGaAs heterostructures[5,6] and spin blockade within Si systems[7,8] are recent advances toward that goal. Silicon offers the potential for very long spin decoherence lifetimes in both donors and quantum dots. The Si metal-oxide-semiconductor (MOS) system features highly tunable carrier densities, the possibility of very small dot sizes, and the promise of complementary MOS (CMOS) compatibility,[9-11] as well as providing further opportunities to couple single donors to gated quantum dots.[4]

Readout of electron spin qubits in quantum dots or donors frequently requires charge sensing[12] to measure the spin state of the qubit by spin to charge conversion.[13-15] Charge sensing, furthermore, is an invaluable tool for achieving few-electron quantum dot occupation in experiments for which transport is suppressed by opaque tunnel barriers in and out of the quantum dot.[16,17] Demonstration of laterally coupled charge sensing in a MOS system is therefore a critical step toward examining the viability of MOS quantum dot and donor-based quantum computing architectures.

Here we report transport measurements through a lateral enhancement-mode quantum dot fabricated using a double-layer MOS gate stack. Modulation of the current through a charge sensing constriction is used to detect single electron changes in the occupation of a neighboring quantum dot. Charge transitions are observed with the charge sensor for both low and high conductance through the quantum dot, including the case when transport through the quantum dot is below the noise floor of the experiment. Both the charge sensing constriction and the quantum dot are stable over long times, and the charge sensing signal observed in this MOS double top gated geometry is comparable to those reported for depletion mode (i.e., modulation doped) GaAs/AlGaAs and Si/SiGe structures.[12,16,18] The charge sensing signal is found to be consistent with measured dot capacitances and predictions from a 3D capacitance model that accounts for the complex topography of the top metal inversion gate. We note that the structure both has an open lateral geometry and uses poly-silicon for the first level of gates, making this device structure compatible with a future capability for self-aligned single donor implantation near quantum dots and charge sensors.[19,20]

The structure studied here was fabricated on a lightly doped p-type silicon wafer (2-20 ohm-cm). Source and drain regions were formed by implantation with arsenic followed by an activation anneal and thermal oxidation to form a 35nm $SiO_2$ gate oxide. The patterned depletion gates immediately above the gate oxide were formed from degenerately doped polysilicon and are shown in Fig. 1a. The entire device was exposed to a second thermal oxidation resulting in 30 nm of $SiO_2$ on the polysilicon depletion gates (and a slight increase in thickness of the exposed gate oxide), followed by atomic layer deposition (ALD) of 60nm of $Al_2O_3$. A metallic top-gate covers the sample and is used to induce-carriers at the $Si/SiO_2$ interface. A cross-sectional schematic of the quantum dot and charge sensing areas of the device is shown in Fig. 1b. Details of the fabrication process flow can be found in Ref. 11. In this work we focus on the gates highlighted in yellow, so that the device operates as a single dot coupled to a charge sensing constriction. All electrical transport measurements were performed in a dilution refrigerator operating at or below a temperature of 150mK. In regions away from the patterned quantum dot, this device functions as a field-effect transistor, and the peak mobility was measured to be 4500 cm$^2$/Vs.

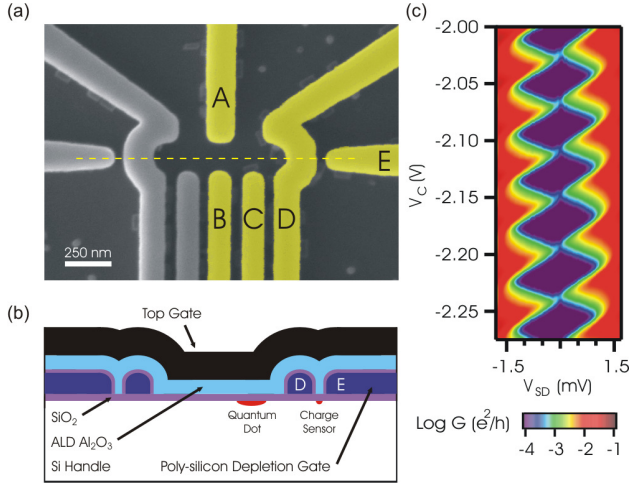

**Figure 1** (a) Scanning electron microscope (SEM) image of a Si nanostructure before deposition of the secondary dielectric and the Al top gate. Gates used for the single dot experiments discussed here are highlighted in yellow. (b) A cross-sectional schematic of the device shown in (a) after completion of the fabrication process. Approximate quantum dot locations are noted. (c) Conductance through the quantum dot as a function of the source-drain voltage $V_{SD}$ and the voltage on gate C. $V_{Top\ Gate}$ = +5 V, $V_A$ = -900 mV, $V_B$ = -500 mV, $V_D$ = -2.1 V, and $V_E$ = 0V.

Fig. 1c shows the differential conductance through the quantum dot as a function of the DC source-drain voltage $V_{SD}$ and the voltage applied to gate C. The differential conductance was measured using a lock-in technique with an AC voltage (50μV at 13Hz) added to $V_{SD}$. The Coulomb diamonds shown indicate the charging energy of the quantum dot is $E_C \approx 1.1$meV. The positions of the Coulomb blockade peaks in this regime were quite stable and displayed an average drift of 0.4% of the oscillation period per day. Capacitance simulations of the device confirm that the Coulomb diamonds shown in Fig. 1c correspond to a quantum dot whose confinement potential is dominated by the electrostatic potential from the lithographic gates highlighted in Fig. 1a.[11]

In contrast to the main quantum dot, the neighboring charge sensing channel operates in the subthreshold regime. This difference in behavior arises because the charge sensing constriction is only 50 nm wide, much narrower than the point contacts for the main quantum dot, which are 145 nm wide (Fig. 1a). As a result, the ALD Al$_2$O$_3$ fills the narrower charge sensing constriction completely, pushing the enhancement top gate away from the channel, and leading to considerably lower top gate coupling to the charge sensing channel than to the main quantum dot. The top-gate bias window over which transport can be observed through the constriction is limited on the low end by the resulting small top gate to channel capacitance and on the high end by the onset of conduction under gate E.

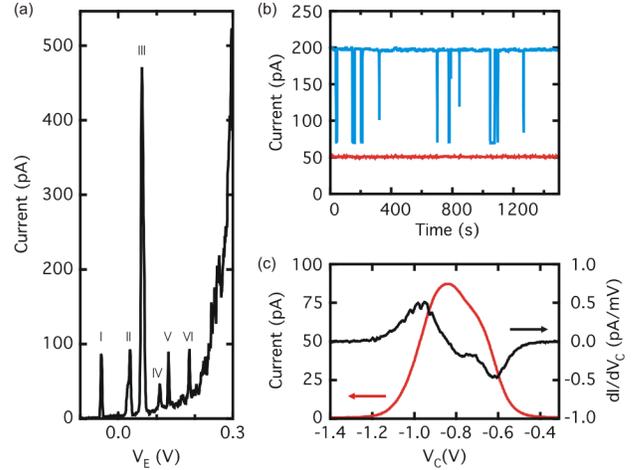

**Figure 2** (a) Current through the charge sensing constriction as a function of the voltage on gate E under conditions where there are no charge transitions in the primary quantum dot. For $V_E$ less than + 0.2 V, the measured current is dominated by transport through the charge sensing constriction and displays Coulomb blockade behavior due to unintentional quantum dot formation. For $V_E$ greater than + 0.2 V, the measured current is not localized near the charge sensing constriction, but rather it is dominated by electron transport under gate E itself. (b) Current through the charge sensing region as a function of time for peak I (red trace) at $V_E$ = -46.7 mV and peak III (blue trace) at $V_E$ = +58.5 mV. Telegraph noise seen in peak III and peak II (not shown) is absent in peak I data. (c) Peak I current as a function of $V_C$ with $V_E$ fixed at -45.5 mV (red trace) along with its numerical derivative (black trace). Charge sensing oscillations are visible in the derivative near $V_C$ = -1 V.

We find that the highest sensitivity to small variations in local potential occurs when the gate voltage $V_E$ is tuned such that disorder dots are formed in the charge sensing channel. Fig. 2a shows Coulomb oscillations through these unintentional quantum dots. Two sets of disorder dot peaks are visible: peaks I, II, III, V, and VI exhibit a similar shift as the top gate voltage is changed slightly (not shown), whereas peak IV behaves quite differently. We therefore believe that peaks I, II, III, V, and VI arise from a single dot. The size of this disorder dot was estimated to be ~25 nm, based on the average Coulomb blockade oscillation period as a function of the overall top gate voltage. This size is also consistent with the lithographic size of the charge sensing constriction Fig. 2b shows the current as a function of time when the operating point is tuned to peaks I or III. As is clear from the data, there is a two-level fluctuator that is active for some gate voltage configurations but not others. Overall, peak I corresponds to the most stable configuration, and all of the data reported below makes use of this peak for charge sensing.

Figure 3 shows charge sensing measurements of the quantum dot. For maximum charge sensitivity, the point contact channel is maintained at the largest slope of peak I by making a compensating voltage step of 13 µV in the voltage on gate E for every 1 mV change in the voltage on gate C. The currents through the QPC constriction and through the quantum dot were recorded sequentially under the same gate bias conditions (black and red curves, respectively). Sharp changes on the charge sensing trace correlate well with the Coulomb blockade conductance oscillations in the quantum dot. Conductance jumps corresponding to charge transitions persist even after Coulomb blockade oscillation amplitude has been reduced below the noise floor of our measurement. Under these conditions, a charge sensing current change of ~2% is observed for each charging event, corresponding to a current change $\Delta I = 0.7$ pA. This fractional change in conduction is similar to other reported results, but occurs here with a substantially lower current, due to the narrow width of the charge sensing constriction used here.[12, 16, 18]

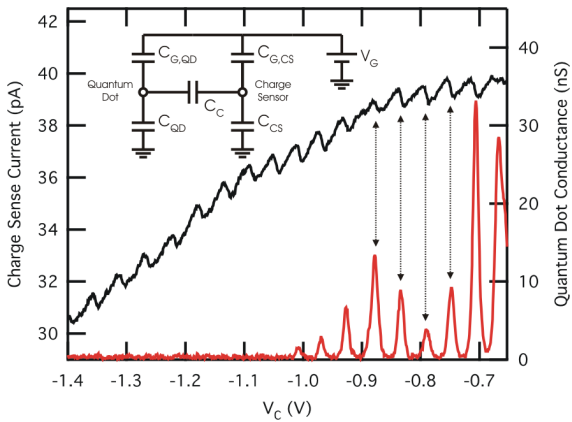

**Figure 3** Current through the charge sensing constriction (black curve), with $V_E$ chosen so that the charge sensor is aligned at the most sensitive region on peak I of Fig. 2(a). Oscillations correspond to charge transitions at the same gate voltages as the Coulomb blockade peaks in the measurement of the quantum dot conductance (red curve). The charge sensor current was measured with an ac voltage of 400µV at 10Hz applied across the QPC ohmic contacts. As $V_C$ was stepped from -0.675 V to -1.4V, $V_E$ was stepped linearly from -42.46mV to -50.57mV in order to compensate for capacitive coupling between the gate C and the charge sensor. Inset: equivalent circuit diagram of the quantum dot and charge sensor system.

Using CFD-ACE+,[21] a capacitive matrix was calculated for the enhancement gate, depletion gates, and quantum dots located in this structure. Gate dimensions were obtained from SEM micrographs. The sizes and positions of the quantum dots, both in the charge sensing channel and within the larger nanostructure, were estimated using the measured capacitances between each dot and the nearby gates.[11] From the simulations, we estimate the coupling capacitance between the main quantum dot and the charge sensing quantum dot to be $C_C = 0.092$ aF.

An effective circuit diagram of the measured device is shown in the inset of Fig. 3. The addition or subtraction of a charge from the quantum dot shifts the potential of the charge sensing dot through the coupling capacitance $C_C$, manifesting as a change in conductance through the charge sensor. Under the assumption that $C_C$ is small compared to the total capacitance of either quantum dot, the voltage change $\Delta V_G$ for a depletion gate that produces a change in the potential on the charge sensing dot equivalent to that due to the addition of an electron to the main dot is given by

$$\Delta V_G = \frac{1}{C_{G,CS}}\left(e\frac{C_C}{C_{\Sigma QD}}\right), \quad (1)$$

where $C_{G,CS}$ is the capacitance between the depletion gate and the charge sensing dot, $C_C$ is the dot-to-dot capacitive coupling, and $C_{\Sigma QD}$ is the total capacitance of the main quantum dot. The balancing voltage $\Delta V_G$ can be found experimentally using $\Delta I = \frac{dI}{dV_G}\Delta V_G$, using the charge sensing signal $\Delta I$ reported above, and the derivative $\frac{dI}{dV_G} = 0.5\,pA/mV$, which is the magnitude of the current change due to capacitive coupling between that gate C and the charge sensing dot at the operating point (Fig. 2c). Using the above relation we find $C_C = 0.09 \pm 0.01$ aF, in good agreement with the calculated interdot capacitance.

Charge sensing sensitivity may be increased in future devices in several ways. First, a wider charge sensing constriction will produce a larger current and a larger signal to noise ratio for the charge sensing measurement. Second, geometric improvements, such as a narrower gate D, or inserting a gap in that gate,[22] will increase the capacitive coupling $C_C$ between the main dot and the charge sensing region and thus will increase the charge sensing signal. Finally, the charge sensor sensitivity also increases as the ratio $C_C/C_{\Sigma QD}$ increases, because this ratio determines the fractional charge induced on the sensor dot due to a single electron change on the main dot.

In summary, integrated charge sensing has been demonstrated in a Si-MOS system by monitoring the current change through a sub-threshold resonance in a nearby conducting channel. Remote charge detection persists into regions of suppressed main dot conductance, an important tool for operation in the few electron regime and for monitoring quantum dot manipulations when conventional quantum dot transport measurements are impractical or impossible. A measure of charge sensing sensitivity in the form of capacitive coupling between the charge sensor and the main dot has been experimentally extracted for the device and shown to agree with a capacitance simulation. This sensitivity can be increased with changes in device geometry.

This work was performed, in part, at the Center for Integrated Nanotechnologies, a U.S. DOE, Office of Basic Energy Sciences user facility. The work at both Sandia National Labs and the University of Wisconsin was supported by the Sandia National Laboratories Directed Research and Development Program. Sandia National Laboratories is a multi-program laboratory operated by Sandia Corporation, a Lockheed-Martin Company, for the U. S. Department of Energy under Contract No. DE-AC04-94AL85000.